\documentclass[journal,12pt,onecolumn,draftclsnofoot,]{IEEEtran}
\usepackage{amsmath,amsfonts}
\usepackage{algorithmic}
\usepackage{algorithm}
\usepackage{array}
\usepackage[caption=false,font=normalsize,labelfont=sf,textfont=sf]{subfig}
\usepackage{textcomp}
\usepackage{stfloats}
\usepackage{url}
\usepackage{verbatim}
\usepackage{graphicx}
\usepackage{cite}
\begin{document}

\title{A Novel Pre-Coding Based PAPR Reduction Scheme for IM/DD Systems}

\author{Idris Cinemre\IEEEmembership{}, and Gokce Hacioglu\IEEEmembership{}% <-this % stops a space
}

% The paper headers
\markboth{Journal of \LaTeX\ Class Files,~Vol.~14, No.~8, August~2015}%
{Shell \MakeLowercase{\textit{et al.}}: Bare Demo of IEEEtran.cls for IEEE Journals}

\maketitle

\begin{abstract}
Orthogonal frequency division multiplexing (OFDM) is critical for high-speed visible light communication (VLC) transmission; however, it suffers from a high peak-to-average power ratio (PAPR) problem. Among PAPR reduction techniques, pre-coding methods have shown promising advantages such as signal independence and no requirement for signaling overhead. In this paper, we present a novel pre-coding method that combines discrete cosine transform (DCT) and discrete sine transform (DST) for VLC systems using OFDM. Our proposed approach, tailored specifically for intensity modulation and direct detection (IM/DD) systems, aims to further improve the PAPR reduction while preserving bit error rate (BER) performance. Simulation results demonstrate that the proposed method achieves more than a $6$ dB reduction in PAPR compared to the DCT pre-coded method without significant degradation in BER performance.
\end{abstract}

% Note that keywords are not normally used for peerreview papers.

\IEEEpeerreviewmaketitle

\section{Introduction}

\IEEEPARstart{V}{isible} light communication (VLC) is a sophisticated optical wireless communication method that leverages visible light emitted by commercial-off-the-shelf light-emitting diode (LED) sources to transmit data over short distances, modulated by intensity modulation (IM) in a way that satisfies lighting requirements and captured by a photodiode on the receiver end via direct detection (DD). A significant challenge in attaining high data rate communication through VLC is the necessity of being able to effectively exploit the limited spectral bandwidth of white LEDs \cite{4926203}. It is highlighted in the literature that implementing orthogonal frequency division multiplexing (OFDM) as a modulation technique characterized by its high spectral efficiency can offer a viable solution to this challenge. Moreover, OFDM is a compelling choice for high-speed data transmission in VLC, owing to its resilience against inter-symbol interference (ISI), improved optical power efficiency, and adaptability to diverse channel conditions, requiring just a simple one-tap equalizer at the receiver \cite{6415964}. Conventional OFDM, based on fast fourier transform (FFT), generates bipolar and complex symbols unsuitable for VLC, requiring real and unipolar symbols. Hermitian symmetry is employed to produce real-valued symbols, and the two most widespread techniques have been reported to ensure  unipolar and positive symbols: Direct current biased optical OFDM (DCO-OFDM) \cite{6415964}, adding positive DC bias, and asymmetrically clipped optical OFDM (ACO-OFDM) \cite{armstrong2006power}, clipping negative symbols at zero.

A notable drawback of utilizing OFDM in wireless communication is having a high peak-to-average power ratio (PAPR), which arises from the superposition of several distinct orthogonal subcarriers \cite{4446229, 1421929}. In VLC, a high PAPR can cause the LED to operate beyond its designed power levels (i.e in a nonlinear zone), resulting in energy inefficiency, a shorter lifespan of LED, and signal distortion in the transmitted signal which leads to a worsened bit error rate (BER) \cite{inan2009impact}. 
%Also, it lowers the effectiveness of light-to-communication conversion \cite{6638694}. 
Therefore, PAPR reduction methods are essential for VLC systems to retain signal quality and enhance energy efficiency. 
% this non-linear behaviour of the LED can cause signal distortion in the transmitted signal, leading to an increase in the bit error rate (BER).

The PAPR reduction strategies for optical OFDM have been thoroughly investigated, which may be broadly classified into three distinct classes; multiple signalling techniques, signal distortion-based methods, and pre-coding schemes. In multiple signalling approaches, several candidate signals with identical information are generated, and the one with the lowest PAPR is selected for transmission. Selective mapping (SLM) \cite{xiao2015papr}, partial transmit sequence (PTS) \cite{tan2015modified}, pilot-assisted (PA)\cite{6731571}, and tone reservation (TR) \cite{bai2017papr} can be listed as the common methods. These methods typically necessitate supplementary information alongside the actual data, leading to decreased bandwidth efficiency and increased computational complexity. 

The methods based on signal distortion reduce the PAPR by employing techniques such as clipping high signal peaks \cite{6887289}, compressing large peaks utilizing non-linear companding \cite{7784728}, and stretching the constellation \cite{7039193}. Although these solutions are straightforward to implement, they introduce distortion noises that can adversely affect the performance of the system. It should be noted that multiple signalling approaches are frequency domain techniques, whereas distortion methods are time domain techniques applied after the inverse FFT (IFFT) to reduce PAPR.
% It should be noted that multiple signalling approaches such as SLM and PTS are frequency domain techniques, while distortion methods, such as clipping, companding, and peak windowing, are time domain techniques applied after the inverse FFT (IFFT) to mitigate PAPR by distorting the signal. 
%In this approach, mainly, before the IFFT block in the OFDM structure, the modulated data is multiplied by a pre-coding matrix, and after the FFT block, the inverse pre-coding matrix is applied to recover the modulated data. 

The pre-coding PAPR reduction schemes are also implemented in the frequency domain with lower complexity than multiple signalling techniques. In this approach, the modulated data is multiplied by a pre-coding matrix at the transmitter before the IFFT block in the OFDM structure, and the inverse pre-coding matrix is applied at the receiver after the FFT block to recover the modulated data. Various transforms have been employed as pre-coding matrices to reduce PAPR, including discrete cosine transform (DCT), zadoff-chu transform (ZCT), discrete Fourier transform (DFT) \cite{ranjha2013precoding}, walsh-hadamard transform (WHT) \cite{salama2022papr}, discrete hartley transform (DHT) \cite{7103295}, and vandermonde-like matrix (VLM) transform \cite{sharifi2019papr}. Pre-coding has a number of advantages over other approaches, including signal independence, no requirement for signalling overhead, and no significant degradation in BER performance\cite{4138047}.

The inherent energy compaction property of the DCT renders it an appealing technique for pre-coding, as this feature has been successfully employed in various signal processing applications, including image processing and video compression \cite{9023199}.  The literature contains several attempts to further enhance PAPR by combining DCT-precoding with other methods. For instance, the article \cite{taha2022reduced} proposes a hybrid PAPR reduction system for DCO-OFDM, which involves the application of DCT pre-coding, followed by multiplication of the IFFT output with a Gaussian matrix to reduce PAPR. This combined method achieves a modest PAPR improvement of approximately $1$ dB without compromising BER performance when compared to the DCT pre-coded method. In \cite{salama2022papr}, the DCT pre-coding scheme is integrated with the clipping method to decrease PAPR for DCO-OFDM, yielding an enhancement of approximately $2$ dB.
% Furthermore, in \cite{wang2014combining}, clipping is employed alongside DCT pre-coding in an IM/DD system, with the analysis exploring the implications of different clipping ratios. 
The study outlined in \cite{wang2013grouped} simultaneously applies multiple DCT pre-coding techniques to an IM/DD system by grouping subcarriers. Although the PAPR improvement achieved is less than 1 dB, it is noteworthy that using groups of two can lead to a $15\%$ reduction in complexity. The study demonstrates a trade-off between BER and PAPR performance, where it is feasible to reduce PAPR but at the expense of significantly degraded BER performance. In \cite{el2022hybrid}, to minimise PAPR in DCO-OFDM, the paper recommends combining DCT precoding with two distinct companding approaches (specifically, $\mu$-law and A-law). The study reveals around a $3$ dB decrease in PAPR for both methods without a loss in BER performance.

This paper introduces a novel pre-coding method which combines the DCT and discrete sine transform (DST) to create new pre-coding matrices and utilizes the method defined in \cite{hacioglu2021optical}. This approach offers all the advantages of pre-coding PAPR reduction techniques while ensuring efficient PAPR reduction without sacrificing BER performance. The remainder of this paper is organized as follows: Section II introduces the system model of conventional DCO-OFDM, DCT pre-coded DCO-OFDM, and the proposed pre-coded method applied to DCO-OFDM; Section III defines the simulation environment, including the model room and channel response of a typical indoor setting, as well as the PAPR and BER comparison results; and Section IV provides concluding remarks.

%PAPR is a vital measurement tool that should be taken into consideration since a high PAPR value reduces the performance of communication systems. 
\textit{Notations:} The bold small letters are denoted as vectors and bold uppercase letters to refer to matrices. $\left[\cdot\right]^{T}$ represents the transpose of a vector or a matrix, $ z^{*}$ indicates the complex conjugate of $z$, and $E\left[\cdot\right]$ refers to expectation value.

\section{System Model}
\label{sec:systemmodel}
\subsection{FFT-Based DCO-OFDM}
\begin{figure}[h!]
	\centering
	\includegraphics[width=5in]{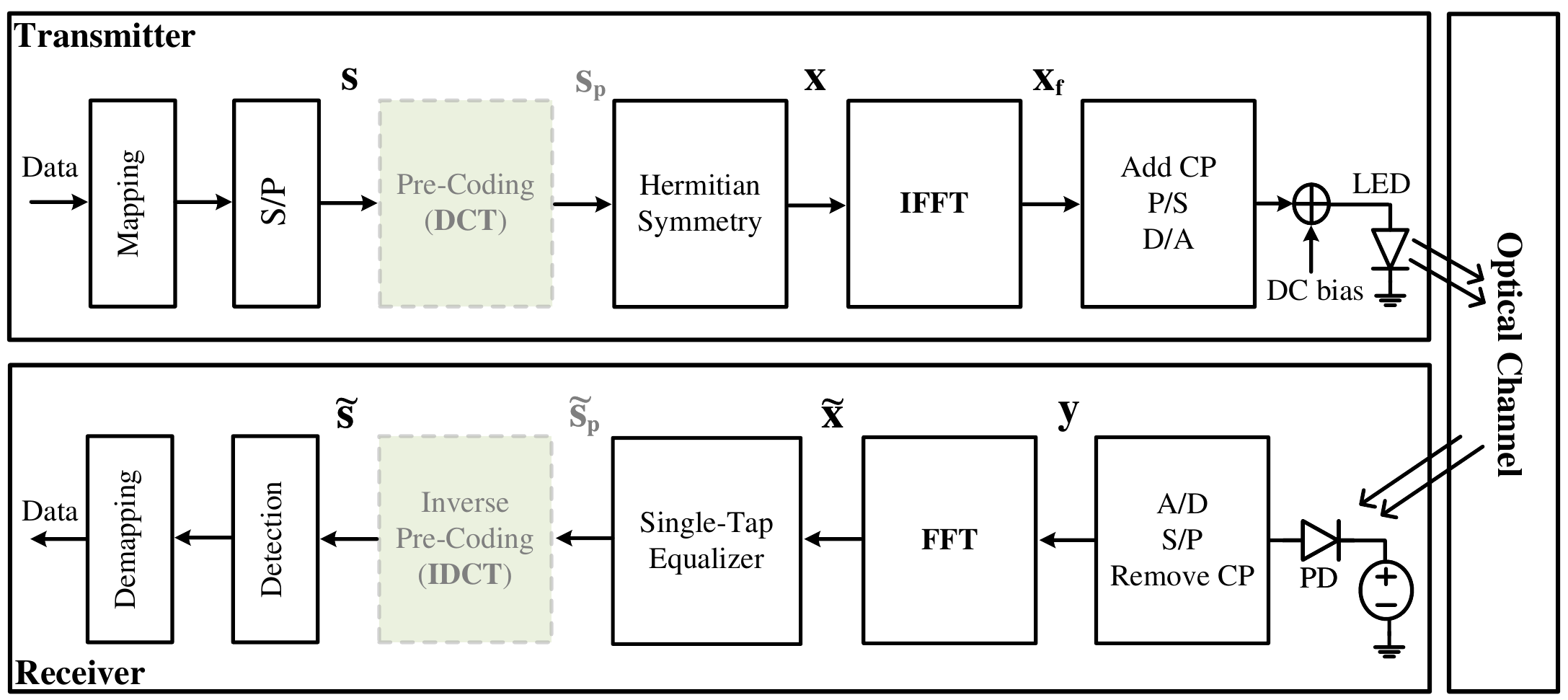}
	\caption{Block diagram of (a) conventional DCO-OFDM (without green boxes) and (b) DCT pre-coded DCO-OFDM(with green boxes)}
	\label{f:DFT_DCO}
\end{figure}

The conventional FFT-based DCO-OFDM system is defined in \cite{6415964}  and given in Fig. \ref{f:DFT_DCO}(a). In DCO-OFDM, data are first mapped into complex-valued symbols ($s_n; n=0,1,\dots,N-1$) based on the modulation scheme (e.g. PSK or QAM). These symbols are stored in $M$-length $\mathbf{x}$ array;  
\begin{align}\label{d: xd}
	&\mathbf{x}=\left[0, s_0, s_1, \ldots, s_{N-1}, 0, s_{N-1}^{*}, \ldots, s_0^{*}, s_1^{*} \right]^{T}, 
	\end{align}
which has to show Hermitian symmetry property to ensure real-valued output, $\mathbf{x_f}$, after IFFT. The IFFT transformation matrix is defined as $\mathbf{IFFT} = \left( f_{k,l}\right)_ {0 \leq k,l \leq M-1 }$ where $f_{k,l}$ is the $\left(k,l\right)^{th}$ element of $\mathbf{IFFT}$ with the size of $M \times M$. 
\begin{equation} \label{d: xf}
	\mathbf{x_f}=\mathbf{IFFT} \times \mathbf{x}
\end{equation}
\begin{equation} \label{d: IFFT}
f_{k,l}=( 1/\sqrt{M} ) e^{\left( j2\pi kl /{M} \right)}
\end{equation}
\noindent It is assumed that the cyclic prefix (CP) with the length of being set to no less than the tap number of the channel impulse response (CIR) is appended prior to transmission. A $\vartheta$-tap CIR is denoted as 
\begin{equation}\label{d: varthetax1}
	\mathbf{h}=\left[h_0, h_1, h_2, \ldots, h_{\vartheta-1}\right].
\end{equation}
\noindent The samples are subsequently formed into a serial sequence and sent to digital-to-analogue (D/A). In DCO-OFDM, a DC bias is then added to the converted electrical signal to eliminate the bipolar nature of the signal and make IM/DD feasible. Note that any remaining negative peaks are required to be clipped before driving the LED to convert this electrical signal into an optical signal.

After passing through the optical channel, the transmitted visible light signal is captured by a photodetector (PD) and subsequently converted to an analogue electrical signal at the receiver; then, an analogue-to-digital converter (A/D) is used to convert the signal to digital, the cyclic prefix (CP) is removed from the samples, and the data is transformed from serial to parallel (S/P). The system operates under the assumption that the channel response is accurately characterized with the $\vartheta$-taps. The received $M$ samples are denoted by $\mathbf{y}$ as in \eqref{d: yrx} where $\mathbf{g}$ stands for additive white Gaussian noise (AWGN) and has $M$ independent and identically distributed (i.i.d) noise terms ($\mathbf{g}=\left[g_0, g_1, g_2,\ldots,g_{M-1}\right]^{T}$). 

\begin{equation} \label{d: yrx}
	\mathbf{y}=\mathbf{C} \times \mathbf{x_f}+\mathbf{g}
\end{equation}
where $\mathbf{C}$ is the circulant channel matrix as defined in \cite{9440940} with the size of $M \times M$. The elements of $\Tilde{\mathbf{x}}$ can be estimated by multiplying $\Tilde{\mathbf{y}}$ with $\mathbf{FFT}$ matrix as
\begin{equation} \label{d: tildex}
\small
\Tilde{\mathbf{x}}=\mathbf{FFT}\times \mathbf{y}=\mathbf{CT}\times \mathbf{x}+\mathbf{FFT}\times \mathbf{g}
\normalsize
\end{equation}
\noindent where $\mathbf{CT}$ is $M \times M$ diagonal matrix given as
\begin{IEEEeqnarray}{lll}
	\mathbf{CT}&=&\mathbf{FFT} \times \mathbf{C} \times\mathbf{IFFT}
	\label{d: CCTNxN}
\end{IEEEeqnarray}
and $\mathbf{FFT}=\mathbf{IFFT}^{H}$.

Unbiased estimations of $\mathbf{s}$ which is placed in $\mathbf{x}$ in \eqref{d: xd}, is denoted as  $\Tilde{\mathbf{s}}$. Because of zero mean additive white gaussian noise (AWGN), the expected value of $E\left[\Tilde{\mathbf{s}}\right]$ is equal to $\mathbf{s}$. The $n^{th}$ elements of  $\Tilde{\mathbf{s}}$ is calculated according to \eqref{d: shat};
\begin{equation} \label{d: shat}
	\Tilde{s}\left({n}\right)=\frac{\Tilde{{x}}\left(n\right)}{CT\left(n, n\right)} 
\end{equation}
where  $n= 0,1,2,\ldots ,N-1$. The estimation of the transmitted $N$ symbols is stored in the $\Tilde{\mathbf{s}}$ array. The receiver makes the decision about $n^{th}$ transmitted symbol $s_n$ by using the maximum likelihood decision (MLD) rule as 
\begin{equation} \label{d: xdet}
\hat{s}_{n}=arg\min_{p_{i}}(\mid \Tilde{s}_{n}-p_{i}\mid^{2})
\end{equation}
where $p_{i}$ denotes one of the $P$ constellation points (i.e. $i=0,1,\ldots,P-1$) and $ \hat{s}_{n}$ denotes $n^{th}$ element of the detected symbols.

\subsection{FFT-Based DCO-OFDM with DCT Pre-Coding}
On the side of the transmitter, pre-coding is employed prior to storing mapped symbols, $\mathbf{s}=\left[ s_0, s_1, \ldots, s_{N-1}\right]^{T}$, into $\mathbf{x}$ in order to achieve a Hermitian property as in Fig. \ref{f:DFT_DCO}(b). The pre-coding matrix is defined as $\mathbf{DCT}=\left( DCT_{k,l}\right)_ {0 \leq k,l \leq N-1}$ where $DCT_{k,l}$ is the $\left(k,l\right)^{th}$ element of $N\times N$ $\mathbf{DCT}$ and given as 
 \begin{equation}
DCT_{k,l}=\begin{cases}
\sqrt{\frac{1}{N}}, & k=0 \\ 
\sqrt{\frac{1}{N}} \cos \left[ \frac{\pi\left(2l+1\right)k}{2N}\right], & k=1, \dots, N-1.
\end{cases}
\end{equation}\label{d: DCTpre}
\noindent The pre-coded symbols, $\mathbf{s_p}$, given in \eqref{d: sp} 
\begin{equation} \label{d: sp}
	\mathbf{s_p}=\mathbf{DCT} \times \mathbf{s}
\end{equation}
are placed into $M$-length $\mathbf{x}$ as in \eqref{d: xpre}
\begin{align}\label{d: xpre}
	&\mathbf{x}=\left[0, \mathbf{s_p}^T, 0, \left( \mathbf{K} \times \mathbf{s_p}^{*} \right)^T \right]^{T}
	\end{align}
where $\mathbf{K}$ is an $N \times N$ exchange matrix such that the elements on the antidiagonal are 1 and the others are 0. After IFFT transform, it follows the same procedure explained in section II-A. 

The inverse pre-coding takes place after equalization by utilizing $\mathbf{IDCT}=\mathbf{DCT}^T$ matrix on the receiver side. 
  \begin{equation} \label{d: Sr2Nx1}
	\mathbf{\Tilde{s}_{p}}=\mathbf{IDCT} \times \mathbf{\Tilde{x}}
\end{equation}
 where  $\mathbf{\Tilde{x}}$ is the output of the FFT transform.  The equalization with the substitution of $\Tilde{s}_{p}\left(k\right)$ in place of $\Tilde{{x}}\left(k\right)$ in \eqref{d: shat} and the detection after pre-coding is implemented using the same approach outlined in the preceding section.
 
\subsection{Proposed Pre-Coded DCO-OFDM}
\begin{figure}[t]
	\centering
	\includegraphics[width=5in]{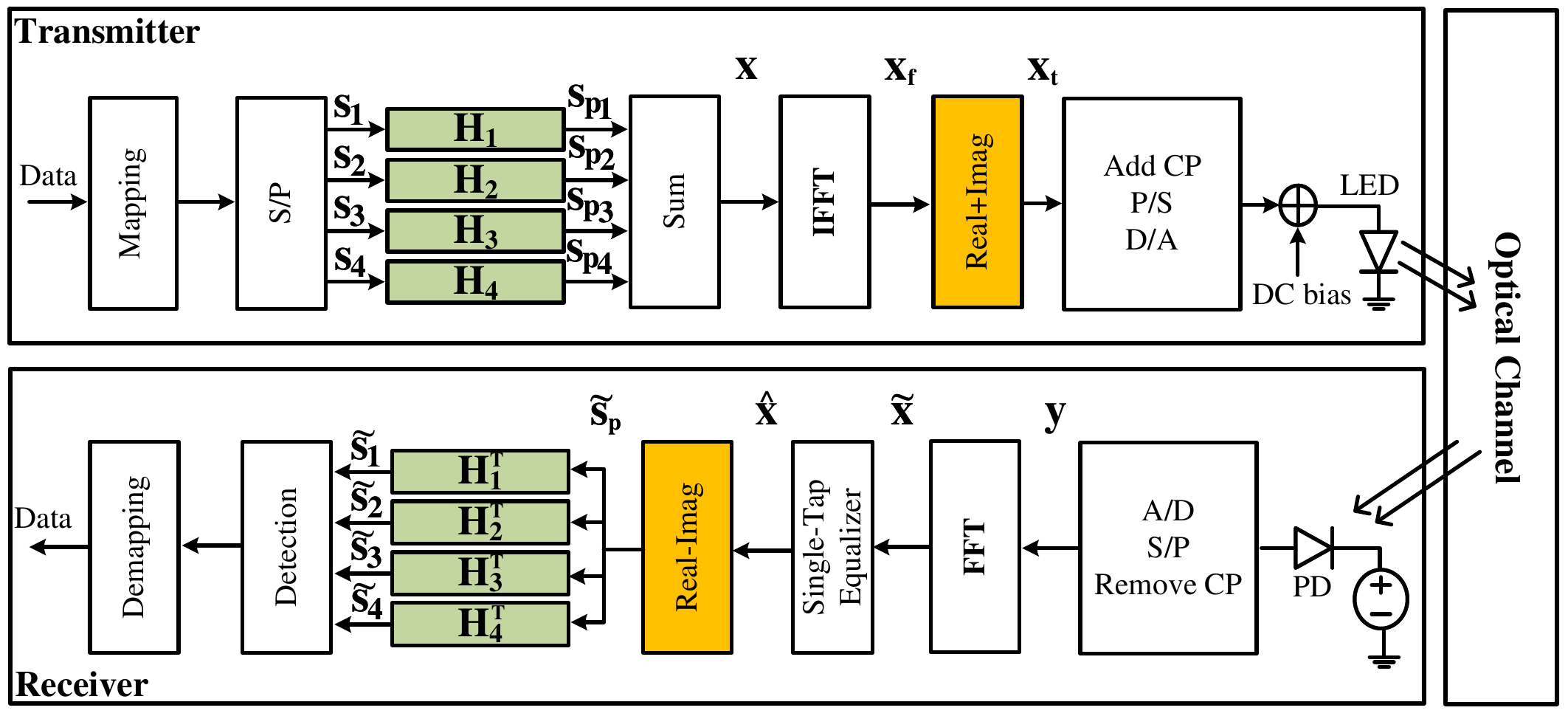}
	\caption{Proposed pre-coded DCO-OFDM method}
	\label{f:proposedblock}
\end{figure}

The proposed DCO-OFDM with the novel pre-coding scheme is given in Fig. \ref{f:proposedblock}. This method combines the DCT and DST to create new pre-coding matrices and utilizes the method proposed in \cite{hacioglu2021optical} instead of the conventional FFT-based OFDM structure with Hermitian symmetry. Basically, to produce real-valued symbols on the transmitter side in this method, the corresponding real and imaginary components of the complex-valued symbols generated at the output of the IFFT are essentially summed instead of employing Hermitian symmetry at the input of the IFFT. Then these components are subtracted from each other after FFT in the receiver. This process will be explained in detail in the following. 

As shown in Fig. \ref{f:proposedblock}, the data are mapped into $N$ real-valued symbols ($s_n; n=0,1,\dots,N-1$) based on the modulation scheme (e.g. PAM) and stored in $\mathbf{s}=\left[ s_0, s_1, \ldots, s_{N-1} \right]^{T}$ which is then divided into four $N/4$-length $\mathbf{s}_{1}$, $\mathbf{s}_{2}$, $\mathbf{s}_{3}$ and $\mathbf{s}_{3}$ arrays.
\begin{subequations}\label{eq:2}
\begin{align}
\mathbf{s_{1}}&=\left[ s_0, s_1, \ldots, s_{N/4-1}\right]^{{T}}  \label{s1}\\
\mathbf{s_{2}}&=\left[s_{N/4}, s_{N/4+1}, \ldots, s_{N/2-1}\right]^{T} \label{s2}\\
\mathbf{s_{3}}&= \left[s_{N/2}, s_{N/2+1}, \ldots, s_{3N/4-1}\right]^{{T}} \label{s3}\\
\mathbf{s_{4}}&=\left[s_{3N/4}, s_{3N/4+1}, \ldots, s_{N-1}\right]^{T}\label{s4}
\end{align}
\end{subequations}

After dividing the data into four equal-length arrays, each of them is pre-coded by the corresponding proposed pre-coding matrix defined below.  These matrices utilize the $\mathbf{DCT}$ given in \eqref{d: DCTpre} and $\mathbf{DST}=\left( DST_{k,l}\right)_ {0 \leq k,l \leq N-1}$ where $DST_{k,l}$ is the $\left(k,l\right)^{th}$ element of $N\times N$ $\mathbf{DST}$ and given as
\begin{equation} \label{d: DST}
DST_{k,l}=\begin{cases}
	\sqrt{\frac{1}{N}}, & k=N-1 \\
		\sqrt{\frac{1}{N}} \sin \left[ \frac{\pi\left(2l+1\right)\left( k+1\right)}{2N} \right], & k=0, \dots, N-2.
\end{cases}
\end{equation}
The matrix notation of obtaining pre-coding matrices is given below where the size of $N/4 \times N/4$ $\mathbf{DCT}$ and $\mathbf{DST}$ and $N/4 \times N/4$ exchange matrix, $\mathbf{K}$, is used.
\begin{subequations}\label{eq:23}
\begin{align}
\mathbf{HDCT} &= \left[ \begin{array}{c}
			\mathbf{DCT}\\
			\mathbf{K}\times \mathbf{DCT} \end{array} \right] \\
   \mathbf{HDST} &=\left[ \begin{array}{c}\mathbf{DST}\\ 
			-\mathbf{K} \times \mathbf{DST} \end{array} \right]
\end{align}
\end{subequations}
\begin{subequations}\label{eq:24}
\begin{align}
\mathbf{D_1} &= \left[ \begin{array}{c}
			\mathbf{HDCT}\\
		 \mathbf{HDCT}
                 \end{array} \right] && (\stepcounter{equation}\text{\theequation}\stepcounter{equation})
                 &\mathbf{D_3}&= \left[ \begin{array}{c}
			\mathbf{HDCT}\\
		 -\mathbf{HDCT}
                 \end{array} \right]\\
             \mathbf{D_2} &=\left[ \begin{array}{c}
                \mathbf{HDST}\\ 
			 \mathbf{HDST}\\
             \end{array} \right] && (\addtocounter{equation}{-1}\text{\theequation}\stepcounter{equation})
                 & \mathbf{D_4} &=\left[ \begin{array}{c}
                \mathbf{HDST}\\ 
			 -\mathbf{HDST}\\
             \end{array} \right]
\end{align}
\end{subequations}
\begin{subequations}\label{eq:4}
\begin{align}
\mathbf{CCT_1}&=\mathbf{D_1}^T \times \mathbf{C_r} \times \mathbf{C_r}^T\times \mathbf{D_1}\label{d: cct1nx1} \\ 
\mathbf{CCT_2}&=\mathbf{D_2}^T \times \mathbf{C_r} \times \mathbf{C_r}^T\times \mathbf{D_2}\label{d: cct2nx1}\\
\mathbf{CCT_3}&=\mathbf{D_3}^T \times \mathbf{C_r} \times \mathbf{C_r}^T\times \mathbf{D_3} \label{d: cct3nx1} \\ 
\mathbf{CCT_4}&=\mathbf{D_4}^T \times \mathbf{C_r} \times \mathbf{C_r}^T\times \mathbf{D_4}\label{d: cct4nx1}
\end{align}
\end{subequations}
\begin{subequations}\label{eq:6}
\begin{align}
\mathbf{H_1}&= \mathbf{D_1} \times \mathbf{E_{\mathbf{CCT_1}}} && (\stepcounter{equation}\text{\theequation}\stepcounter{equation}) 
& \mathbf{H_3}&= \mathbf{D_3} \times \mathbf{E_{\mathbf{CCT_3}}} \\
\mathbf{H_2} &=\mathbf{D_2} \times \mathbf{E_{\mathbf{CCT_2}}} && (\addtocounter{equation}{-1}\text{\theequation}\stepcounter{equation})
& \mathbf{H_4} &=\mathbf{D_4} \times \mathbf{E_{\mathbf{CCT_4}}} 
\end{align}
\end{subequations}
\noindent where $\mathbf{C_r}$ denotes circulant matrix of random typical indoor VLC channel defined in the next section and  $\mathbf{E_{CCT}}$ represents eigenvalues of $\mathbf{{CCT}}$. Note that $\mathbf{C_r}$  is solely utilized for deriving the new pre-coding matrix and remains independent of the channel.  The new pre-coding matrices, $\mathbf{H_1}$, $\mathbf{H_2}$, $\mathbf{H_3}$, and $\mathbf{H_4}$, is produced in the dimension of $N \times N/4$.

The pre-coded real-valued arrays, $\mathbf{s_{p1}}, \mathbf{s_{p2}}, \mathbf{s_{p3}}, \mathbf{s_{p4}}$ are summed and then the IFFT transform is applied. 
\begin{equation} \label{eq1}
\mathbf{x}  = \mathbf{s_{p1}}+\mathbf{s_{p2}}+\mathbf{s_{p3}}+\mathbf{s_{p4}} \\
\end{equation}
\begin{equation} \label{d: x}
	\mathbf{x_f}=\mathbf{IFFT} \times \left( \mathbf{H_1} \times \mathbf{s_{1}} + \mathbf{H_2} \times \mathbf{s_{2}} +\mathbf{H_3} \times \mathbf{s_{3}} +\mathbf{H_4} \times \mathbf{s_{4}} \right)
\end{equation}

The complex output of the IFFT is converted to a real number in this model by adding the real and imaginary values of each symbol.

\begin{equation} \label{d: xt}
	\mathbf{x_t}= real(\mathbf{x_f}) + imag(\mathbf{x_f})
\end{equation}
Thereafter, the same series of steps explained in section II-A is followed.
 
In receiver side, $\mathbf{y}$ is calculated as same the equation in \eqref{d: yrx} by replacing $\mathbf{x_f}$ with $\mathbf{x_t}$. The elements of $\Tilde{\mathbf{x}}$ can be calculated as in \eqref{d: tildex} where $\mathbf{CT}$ is $N \times N$ diagonal matrix defined in  \eqref{d: CCTNxN}.
After equalization as in \eqref{d: xfhat}, 
\begin{equation} \label{d: xfhat}
	\hat{x}\left({k}\right)=\frac{\Tilde{{x}}\left(k\right)}{CT\left(k, k\right)} 
\end{equation}
the imaginary part of the $\hat{\mathbf{x}}$ is subtracted from the real part of the $\hat{\mathbf{x}}$
\begin{equation} \label{d: spt}
	\mathbf{\Tilde{s}_{p}}= real(\mathbf{\hat{x}}) - imag(\mathbf{\hat{x}}).
\end{equation}

Then, inverse pre-coding is applied to $\mathbf{\Tilde{s}_{p}}$ to obtain respective elements of estimated data arrays, $\mathbf{\Tilde{s}_{1}}$, $\mathbf{\Tilde{s}_{2}}$, $\mathbf{\Tilde{s}_{3}}$, $\mathbf{\Tilde{s}_{4}}$. 
\begin{subequations}\label{eq:25}
\begin{align}
\mathbf{\Tilde{s}_{1}}&= \mathbf{H_1}^T \times \mathbf{\Tilde{s}_{p}} && (\stepcounter{equation}\text{\theequation}\stepcounter{equation}) 
&  \mathbf{\Tilde{s}_{3}}&= \mathbf{H_3}^T \times \mathbf{\Tilde{s}_{p}} \\
\mathbf{\Tilde{s}_{2}} &=\mathbf{H_2}^T \times \mathbf{\Tilde{s}_{p}} && (\addtocounter{equation}{-1}\text{\theequation}\stepcounter{equation})
&  \mathbf{\Tilde{s}_{4}} &=\mathbf{H_4}^T \times \mathbf{\Tilde{s}_{p}}
\end{align}
\end{subequations}
Note that during detection each symbol should be divided into 4 which equals the diagonal elements of $\mathbf{H}^T \times \mathbf{H}$.
\section{Numerical Results}

  \begin{figure}[t]
	\centering
	\includegraphics[scale=0.6]{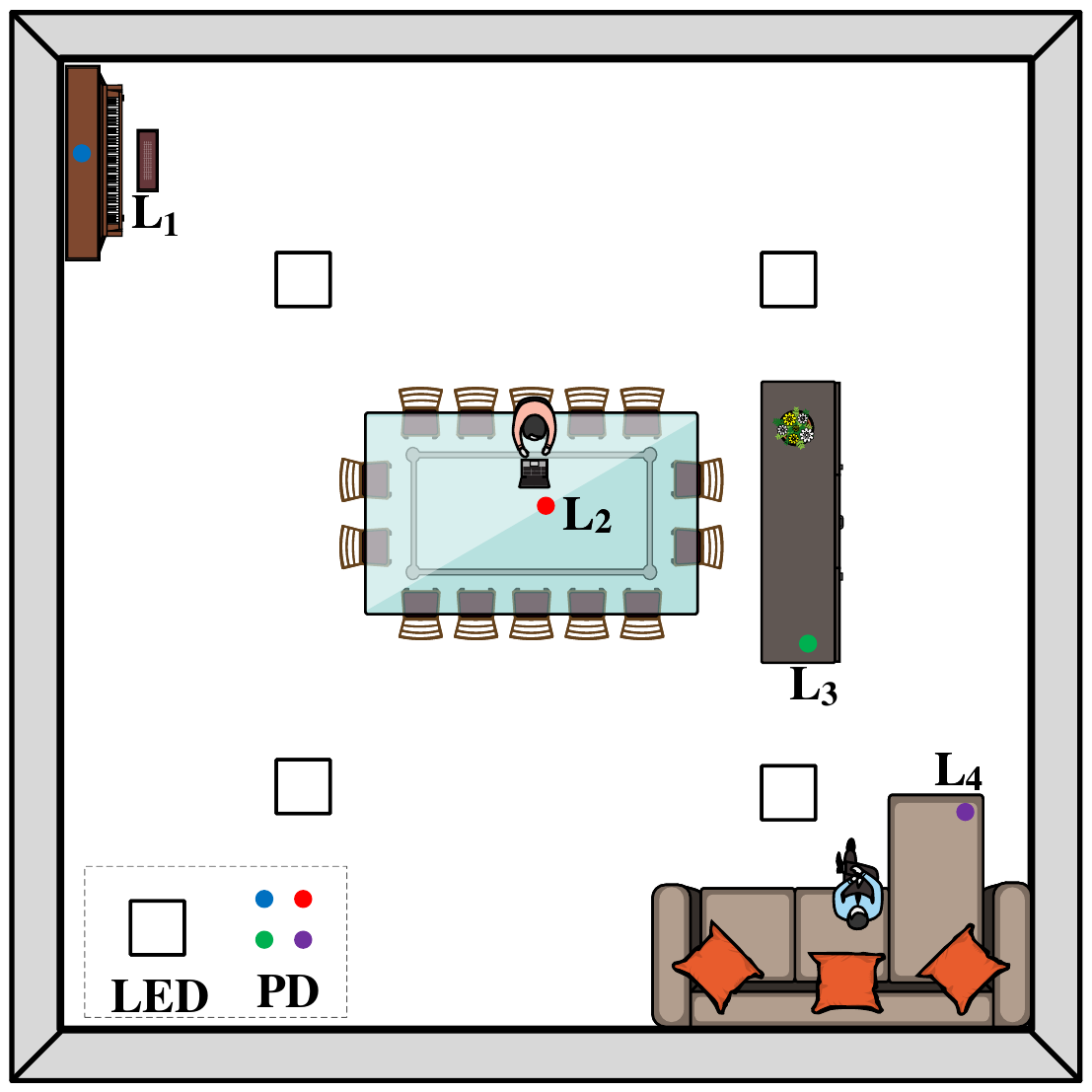}
	\caption{Top view of model room}
	\label{f:room}
\end{figure}

The size of $ 5\times5\times3 m $ model room with four LEDs is considered for simulations. The top view of the model room is shown in Fig. \ref{f:room}. The $(x,y,z)$ coordinates of the luminaries are $\left(1.25,3.75,3m\right)$, $\left(3.75,3.75,3m\right)$, $\left(1.25,1.25,3m \right)$ and  $\left(3.75,1.25,3m\right)$. It is assumed that only the side walls, the ceiling and the floor reflect the light with reflection coefficients of $0.8$, $0.7$ and $0.2$, respectively. Receiver field of view is $85^{\circ}$ and physical area of photo detector (PD) is $10^{-4}m^{2}$ and semi-angle at half power of LED is $55^{\circ}$. The coordinates of PDs are given in the figure. 

\begin{figure}[b]
	\centering
	\includegraphics[scale=0.4]{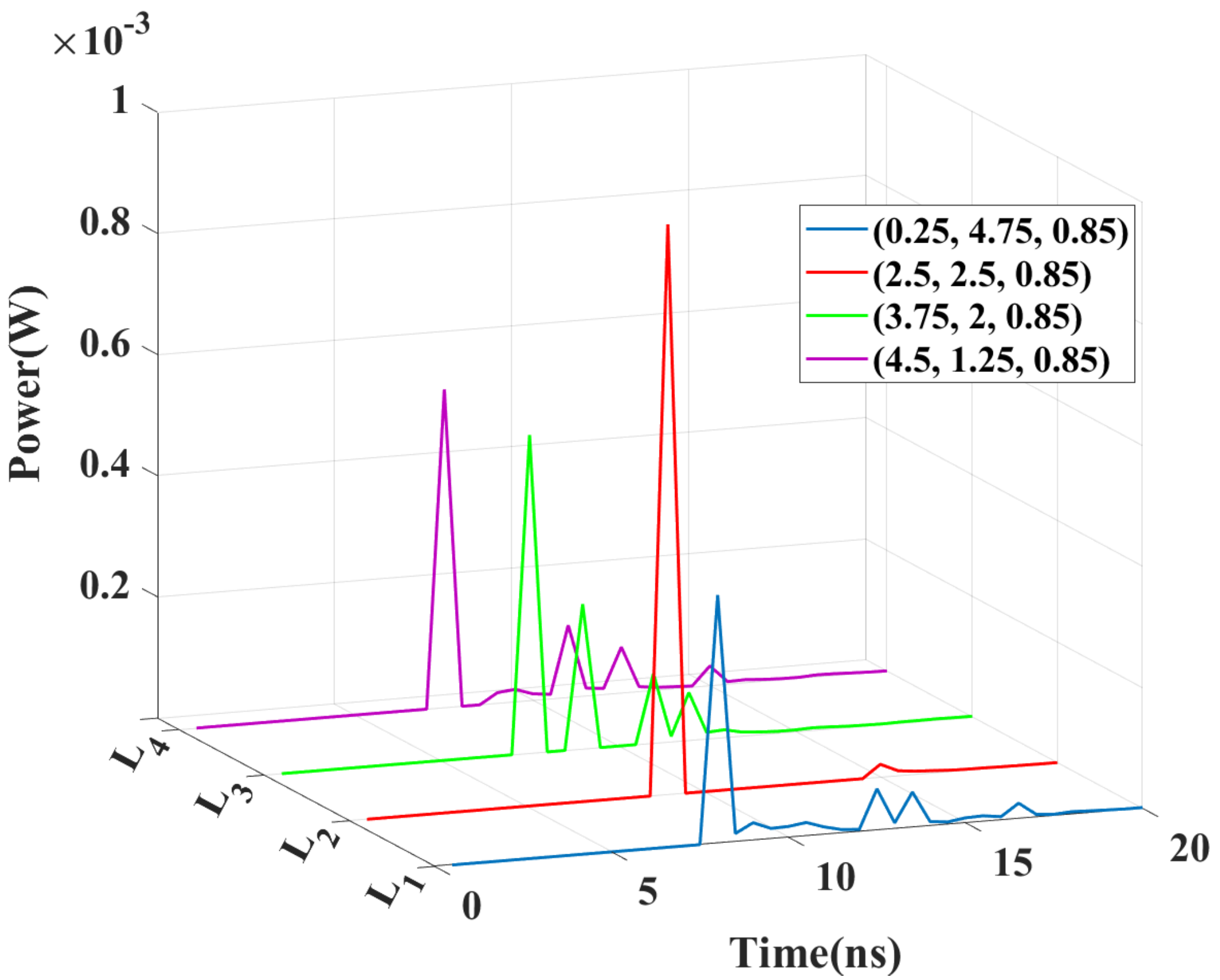}
	\caption{Channel impulse response at the receiver}
	\label{f:CIR}
\end{figure}

The CIRs at various locations $\left(L_1, L_2, L_3, L_4\right)$ are obtained using the ray tracing method introduced in \cite{lee2011indoor}, considering the line of sight (LOS) and two reflections (NLOS), as shown in Fig. \ref{f:CIR}. It is assumed that LEDs have lambertian radiation patterns and reflecting surfaces are lambertian reflector for simplification.  Also, we assumed that the typical indoor channel has a strong LOS component (i.e $h_0$, contains $70-85\%$ of received power) and weak NLOS components (i.e $h_1$, contains $15-25\%$ of received power and $h_2$, contains $5-10\%$ of received power) based on our simulations and  \cite{lee2011indoor, 7823364}. According to this assumption, the $\mathbf{C_r}$ utilized in the derivation of pre-coding matrices is created randomly. 

\begin{figure*}[h!]
	\centering
	\subfloat[][]{\includegraphics[width=3.2in]{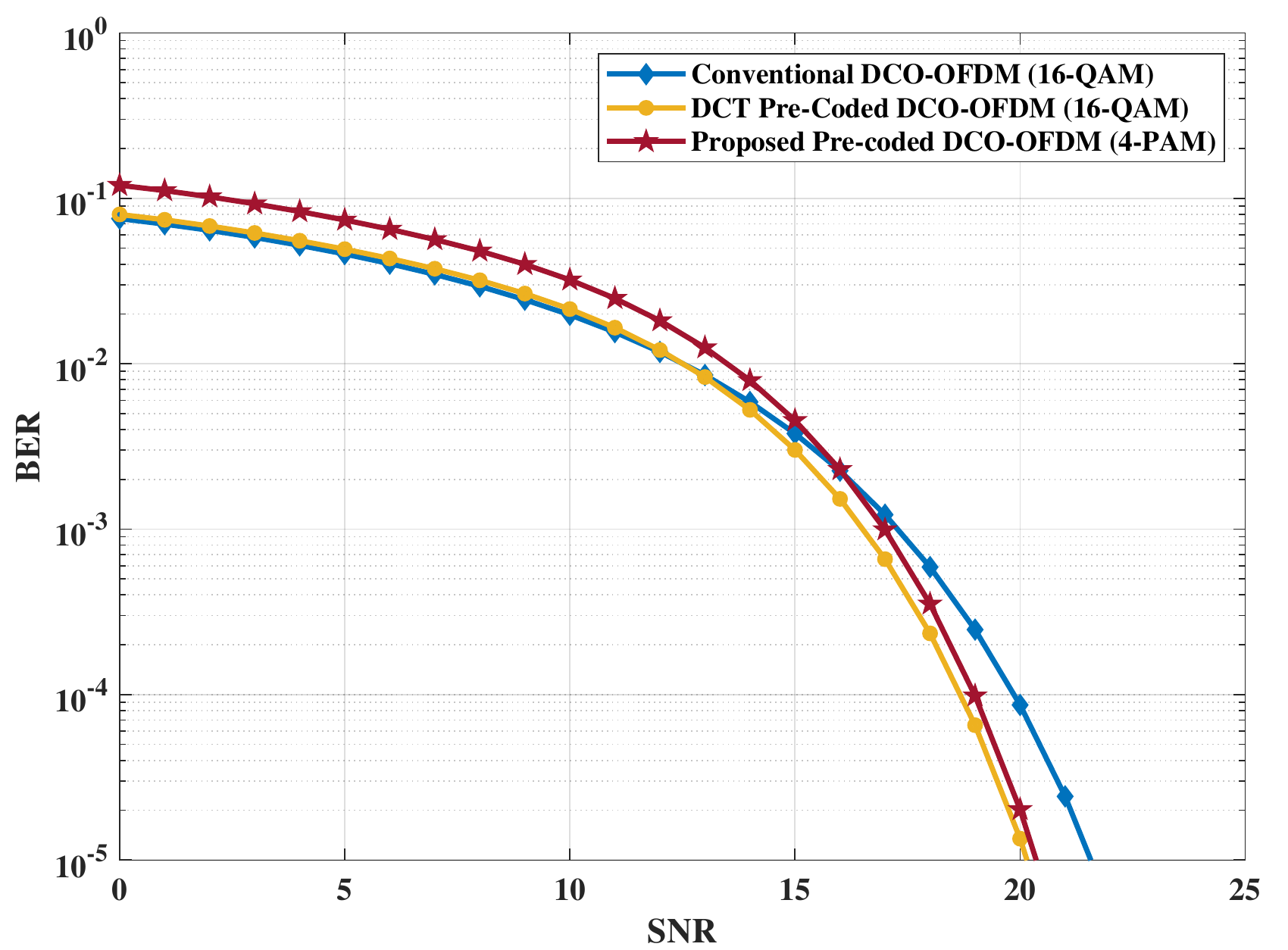}}
		\label{f:bercomp}
	%			\vfill
	\subfloat[][]{\includegraphics[width=3.2in]{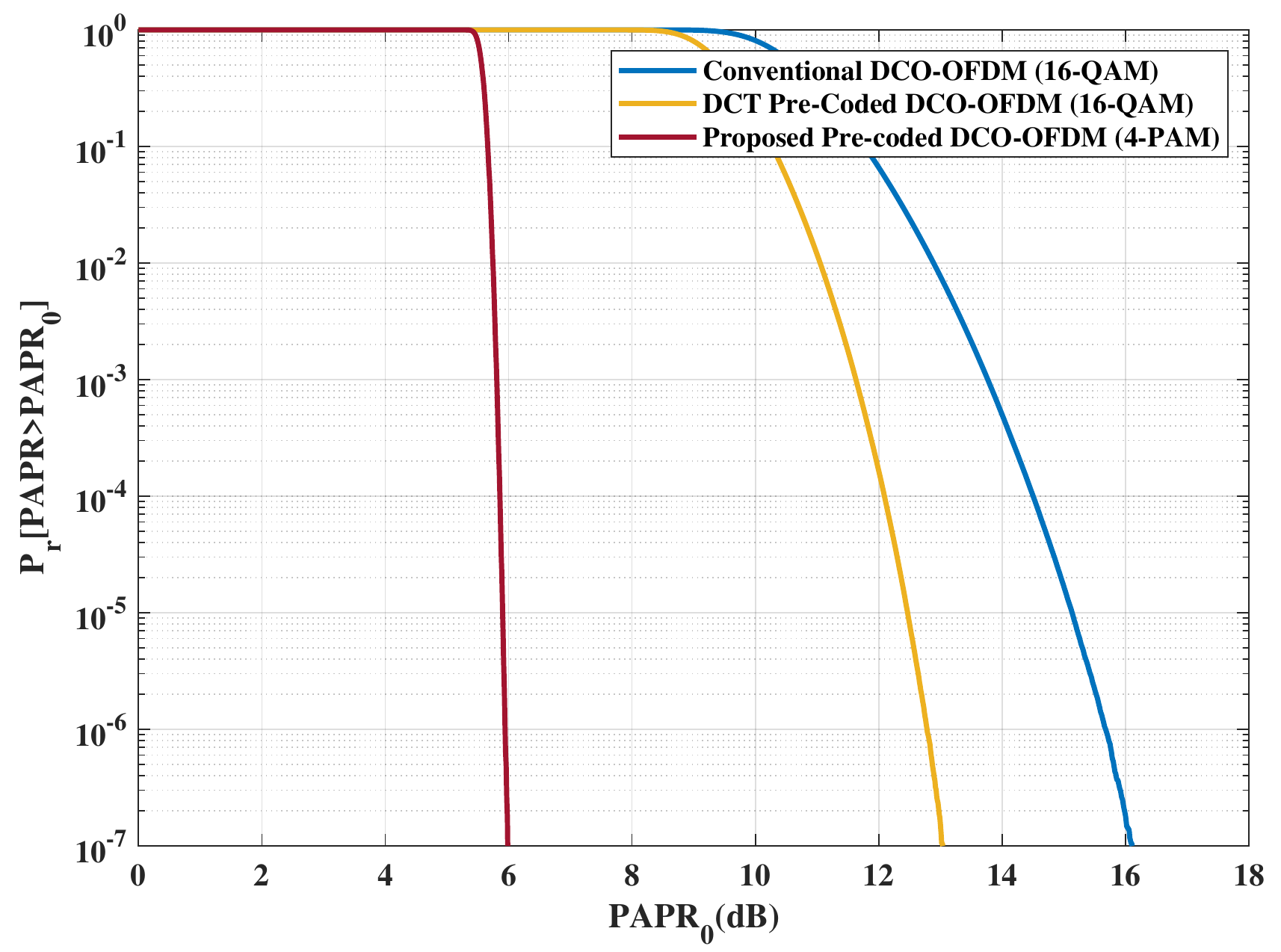}}
		\label{f:paprcomp}
	\caption{ (a) BER and (b) PAPR comparison between the proposed pre-coding and DCT pre-coding DCO-OFDM schemes}
	\label{f:comparison}
\end{figure*}

In the present study, a comparative analysis was conducted to evaluate the performance of BER and PAPR of three distinct DCO-OFDM methods defined in Section II. This evaluation examines the PD placement at location $L_1$ which exhibits a frequency-selective channel response that is used for defining the matrix $\mathbf{C}$. %conventional FFT-based DCO-OFDM, DCT pre-coded FFT-based DCO-OFDM, and FFT-based DCO-OFDM pre-coded with the method proposed in Section II.C. 
To evaluate the PAPR performance of the methods, the distribution of PAPR, commonly expressed through the complementary cumulative distribution function (CCDF) as defined in \cite{9440940}, is utilized. As seen in Fig. \ref{f:comparison}(a), the BER performance of the proposed method, when compared to other methods, does not result in significant degradation. Additionally, the proposed method demonstrates a notably superior PAPR performance, with over $6$ dB reduction compared to the DCT pre-coded method, as shown in Fig. \ref{f:comparison}(b).

\section{Conclusion}

This article proposes a novel pre-coding algorithm used in IM/DD systems. The proposed method involves partitioning the modulated data into four equal subsets and utilizing proposed precoding matrices prior to the IFFT transformation, in conjunction with replacing the Hermitian symmetry through the addition of the real and imaginary components of the symbols prior to transmission. The proposed approach, similar to other methods, employs a single-tap equalizer for IM/DD systems, which guarantees the preservation of BER performance while substantially enhancing PAPR efficiency. The proposed method could allow for the transmission of each subset of the symbols by distinct light sources, and this property may be used for multiple access in future works.

\ifCLASSOPTIONcaptionsoff
\newpage
\fi
\bibliographystyle{IEEEtran}
\bibliography{IEEEabrv,referanslar}

\end{document}